# Shrinking annuli mechanism and stage-dependent rate capability of thin-layer graphite electrodes for lithium-ion batteries

Michael Heß, Petr Novák[1]

Paul Scherrer Institut, Electrochemistry Laboratory, CH-5232 Villigen PSI, Switzerland

**ABSTRACT:** The kinetic performance of graphite particles is difficult to deconvolute from half-cell experiments, where the influences of the working electrode porosity and the counter electrode contribute nonlinearly to the electrochemical response. Therefore, thin-layer electrodes of circa 1 μm thickness were prepared with standard, highly crystalline graphite particles to evaluate their rate capability. The performance was evaluated based on the different stage transitions. We found that the transitions towards the dense stages 1 and 2 with $LiC_6$ in-plane density are one of the main rate limitations for charge and discharge. But surprisingly, the transitions towards the dilute stages 2L, 3L, 4L, and 1L progress very fast and can even compensate for the initial diffusion limitations of the dense stage transitions during discharge. We show the existence of a substantial difference between the diffusion coefficients of the liquid-like stages and the dense stages. We also demonstrate that graphite can be charged at a rate of ~6C (10 min) and discharged at 600C (6 s) while maintaining 80 % of the total specific charge for particles of 3.3 μm median diameter. Based on these findings, we propose a shrinking annuli mechanism which describes the propagation of the different stages in the particle at medium and high rates. Besides the limited applicable overpotential during charge, this mechanism can explain the long-known but as yet unexplained asymmetry between the charge and discharge rate performance of lithium intercalation in graphite.

## 1. Introduction

Graphite is one of the most important negative electrode materials used in current lithium-ion batteries due to its very negative open-circuit potential, improved safety compared to metallic lithium and moderate cost. But it has some major drawbacks, such as its moderate specific charge of 372 mAh/g and its rather low practical rate capability, which is limited to ~2C for lithiation (charge with respect to a complete battery) and 20C for delithiation (discharge) in standard electrodes [1]. However, recent *ab initio* calculation based on density functional theory (DFT) and Monte Carlo simulations [2] yielded a diffusion coefficient larger than $D = 2 \cdot 10^{-8}$ cm$^2$/s for the diffusion of lithium in dense stage 1 and 2 in the presence of many vacancies. But the diffusion coefficient decreases to $5 \cdot 10^{-9}$ cm$^2$/s for fully lithiated stages 1 and 2 due to a low number of vacancies for Li hopping [2]. Similar results have been found experimentally by Levi et al. [3], who determined the diffusion coefficient by applying the potentiometric intermittent titration technique at 60 °C using the Cottrell equation. These experiments revealed highly fluctuating diffusion coefficients of approximately $5 \cdot 10^{-9}$ cm$^2$/s for the transitions involving dense stage 1 and 2 that drop sharply to $10^{-10}$ cm$^2$/s at the end of each phase transition. For the liquid-like phases, i.e., stage 2L, 3L, 4L, and 1L, the diffusion coefficient was determined to be approximately $1-5 \cdot 10^{-8}$ cm$^2$/s, dropping to a minimum of $10^{-9}$ cm$^2$/s at the end of each phase transition[3]. This suggests a difference in the lithium diffusion coefficients of a factor of 10 for the liquid-like stages compared to the dense stages with $LiC_6$ in-plane density. In any case, from both the *ab initio* calculation [2] and the experiment [3], one would expect the rate capability of graphite to be higher than ~2C for charge and 20C for discharge [1] for particles in the micrometer range.

It is known that the performance of graphite electrodes depends significantly on the electrode preparation, due to its influence on the porosity, the tortuosity, and the overall electronic conductivity of the graphite network [1, 4]. To characterize the properties of graphite particles themselves, however, it is necessary to avoid parasitic effects, such as electrolyte diffusion limitation within the porous electrode, electronic conductivity limitations, and the influence of the counter electrode. It has been shown for electrochemical impedance spectroscopy and cyclic voltammetry that thin layers of graphite can be used to separate the different electrochemical influences in a graphite electrode [5]. Therefore, we prepared thin-layer electrodes of graphite flakes in this work and, thus, aimed at avoiding porosity effects and reducing the active mass and consequently the specific currents applied to the electrolyte and the lithium counter electrode. These thin layers consist of 2-3 flat graphite flakes stacked above one another (approximately 1 μm thickness) and directly reflect the particles' properties.

To understand the kinetics of the lithiation and delithiation of graphite, we separated the results into the different stage transitions. The stages of the lithium-graphite binary system at room temperature are depicted in Figure 1. The stages are named by the ordering of the lithium ions in every n$^{th}$ interslab. Additionally, the letter L indicates liquid-like, disordered lithium ordering in every n$^{th}$ interslab. Starting from pure graphite, 4-7 % of lithium with respect to $LiC_6$ (state-of-charge (SOC) of 100 %) can be intercalated into stage 1L (solid-solution of lithium in graphite) [6], where every layer is filled with lithium ions in a liquid-like manner, i.e., without any in-plane order. Then, a phase separation occurs to form stage 4L from stage 1L, with every forth interslab filled in a liquid-like manner. A debate about the transi-tions between stages 4L, 3L, and 2L is still ongoing. X-ray diffraction experiments (XRD) indicate that the c-axis parameter shifts continuously with increasing lithium intercalation [6] without a clear sign of phase separation. By contrast, minor peaks appear in the cyclic voltammetry, indicating stage transitions [5]. However, it is known that stage 3L consists of disordered lithium ions in every third interslab. The next stage transition occurs from stage 2L, with every second interslab filled without any in-plane order, to the ordered dense stage 2, with $LiC_6$ in-plane ordering in every other interslab [7]. Finally, a phase separation from stage 2 to stage 1 occurs, forming a $LiC_6$ structure in every interslab. In dense stages 1 and 2, the lithium ions align in straight columns, unlike all other alkali and earth alkali metals [8]. The intercalation kinetics of graphite can be understood in terms of these stage transitions.

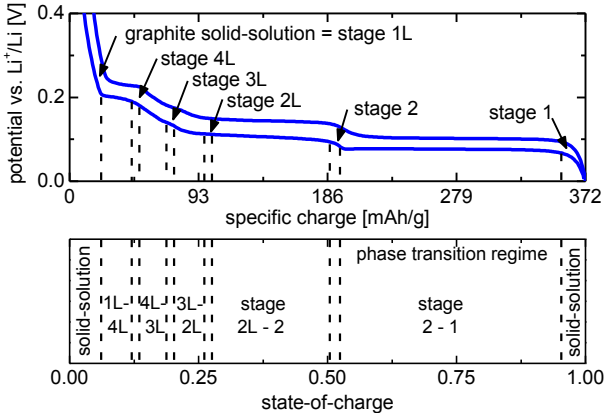

Figure 1: Different stages and stage transitions of graphite during lithium intercalation, dashed lines indicate solid-solution regime of each phase at room temperature with coexistence regime between pure phases.

In the first section of this article, the kinetics of a porous electrode will be discussed based on the work of Doyle et al. [9]. Their equation system will be used to identify the main influences governing the thin-layer half-cell electrode. In the second section, we will use the thin-layer model electrode to investigate the rate capability of graphite considering the side reactions occurring in these electrodes. In the last section, a model is proposed to explain the asymmetry of the rate capability between charge and discharge.

## 2. Experimental

Graphite SFG6 (TIMCAL) was used for all experiments. This sample has a specific surface area of approximately 17.1 m$^2$/g (from BET) and 50 % and 90 % of the graphite particles are smaller than 3.3 and 6 µm along the a and b-axis, respectively [10]. The dimension of the flakes along the c-axis of the SFG6 graphite sample is approximately 0.4 µm, as estimated from the scanning electron microscopy image shown in Figure 2a. This kind of graphite consists of highly crystalline graphite flakes with a primary crystallite size along the a and b-axis of ~30 nm ($L_a$) and ~60 nm along the c-axis ($L_c$), as determined from XRD [10]. Slurries of graphite (90 wt%) and PVDF binder (Kynar Flex, 10 wt%) in N-methyl-pyrrolidone (NMP) were used to cast the electrodes. For thin-layer electrodes, 3 wt% SuperP conductive filler (TIMCAL) were added to give a mass ratio of 87:10:3 to account for the lower in-plane electronic conductivity due to reduction in the number of conduction pathways compared to standard electrodes. The test electrodes were cast onto a copper foil current collector by two different methods. For standard electrodes with thicknesses of more than 30 µm, the doctor blade technique was used. For thin-layer electrodes of less than 20 µm, a spray nozzle technique was employed. The coated copper foil was dried in vacuum at 80 °C in order to remove NMP. The heating time was minimized to reduce the risk of oxidation of the copper current collector. Subsequently, electrodes were punched out (13 mm diameter) and dried under vacuum at 120 °C for 12 h in order to remove remaining water and NMP. The electrochemical coin-type test cells made from titanium were assembled in an argon-filled glove box, under continuous removal of $N_2$, $O_2$, $H_2O$, and organic volatiles. The SuperP electrodes used as a reference were prepared with PVDF (Kynar Flex) in a mass ratio of 80:20 and coated from a NMP slurry. An increased amount of PVDF was chosen due to the small particle size of SuperP carbon. For comparison, blank copper current collectors were coated with pure NMP and dried in a similar manner. 500 µl of 1M $LiPF_6$ in EC:DMC 1:1 (wt%, Novolyte) and a lithium counter electrode with a diameter of 13 mm were used in all experiments.

Electrochemical cycling was performed using Bat-small (Astrol Electronic AG) for standard electrodes, and VMP3 (Biologic) for thin-layer electrodes. The VMP3 was essential because of its fast current response time of 40 µs, which is necessary at a rate of 1000C (3.6 s per half-cycle). 1C corresponds to a charge rate of 372 mA/g. All electrochemical tests were performed in an environmental chamber at 25 ±0.1 °C.

## 3. Theoretical background

Starting from the equation system described by Newman et al. [9, 11] for a porous electrode, the equation system can be simplified considering the implications of the thin-layer electrode. The equations for the potential drop in the electrolyte $\nabla \Phi_e$ and electrolyte diffusion are [9]:

$$\nabla \Phi_e = \frac{-i_e}{\kappa} + \frac{2RT}{F}\left(1-t^0\right)\left(1+\frac{d\ln f_{\pm 0}}{d\ln c_e}\right)\nabla \ln c_e \quad (1)$$

$$\varepsilon \frac{\partial c_e}{\partial t} = \nabla \cdot \varepsilon D_e \left(1 - \frac{d\ln c_0}{d\ln c_e}\right)\nabla c_e + \frac{t_{ne}^0 \nabla \cdot i_e + i_e \cdot \nabla t_{ne}^0}{F} \quad (2)$$

Furthermore, there are four equations governing the electrode kinetics of the negative and positive electrode. The surface reactions can be described by the Butler-Volmer equation (3):

$$I = i_0 e^{\left(\frac{\alpha_a zF}{RT}\eta\right)} - i_0 e^{\left(-\frac{\alpha_c zF}{RT}\eta\right)} \quad (3)$$

$$\eta = \phi_s - \phi_e - U_{OCP} - R_{SEI}i_n \quad (4)$$

The diffusion of lithium in the solid is represented by Fick's second law in equation (5), although one has to keep in mind that graphite undergoes more complicated diffusion than isotropic radial diffusion of an assumed cylinder as equation (5) would imply. Additionally, some of the stage transitions in graphite are phase separating, which requires more sophisticated models, as shown for, e.g., $LiFePO_4$ [12].

$$\frac{\partial c_s}{\partial t} = \frac{1}{r}\frac{\partial}{\partial r}\left(D_s r \frac{\partial c_s}{\partial r}\right) \quad (5)$$

Additionally, the potential drop in the porous electrode matrix $\nabla \phi_s$ can be described by Ohm's law:

$$\nabla \phi_s = \frac{i_e - I}{\sigma} \quad (6)$$

Furthermore, charge conservation is given by:

$$\nabla \cdot i_e = a_i i_n \quad (7)$$

Applying the conditions of a thin layer (monolayer-bilayer) of graphite versus a lithium counter electrode simplifies the equation system of Doyle et al. [9]. By using a lithium counter electrode, one does not need to consider solid-state diffusion and ohmic potential drop in the lithium counter electrode. Additionally, a very thin layer of graphite can be considered equipotential in thickness direction of the electrode (if well connected electronically) neglecting thus the potential distribution in the graphite matrix (eq. 6). Furthermore, the applied geometrical current density I at the current collector is transferred to the thin layer in the matrix as $i_s$ and passes to the electrolyte, $i_e$, via surface reactions. The thin-layer electrode is assumed to suppress any inhomogeneous current distribution in the direction perpendicular to the graphite layer.

To estimate the influence of diffusion of lithium in the active intercalation material and the electrolyte, Doyle et al. [9] defined two dimensionless numbers. They allow the identification of diffusion limitations both in the solid and the electrolyte and give an estimation about the relevance of the particular process. The influence of the solid-state diffusion of lithium is given as [9]:

$$S_s = \frac{r^2 i}{D_s F (1-\varepsilon) c_{s,max} h_s} \tag{8}$$

The dimensionless number of the solid ($S_s$) depends on the ratio of lithium diffusion to the total charge stored where $c_{s,max}$ is the maximum concentration of lithium in the solid active material, $h_s$ the thickness of the porous electrode and $\varepsilon$ the porosity in the electrode. The equivalent dimensionless number to estimate the influence of the diffusion of lithium ions in the electrolyte ($S_e$) was also given by Doyle et al. [9] as:

$$S_e = (h_e + h_s)^2 \frac{i}{D_e F (1-\varepsilon) c_{s,max} h_s} \tag{9}$$

which correlates the diffusion in the electrolyte perpendicular to the graphite layer in the porous electrode ($h_s$) and in the separator ($h_e$) to the total charge stored.

For an electrode containing 0.2 mg of graphite per 1.33 cm$^2$ geometrical area, the current density of 20 A/m$^2$ corresponds to a rate of 36C. Based on the values listed in Table 1, the dimensionless numbers $S_s$ and $S_e$ are 0.28 and 0.29, respectively. The order of magnitude of the dimensionless numbers being close to 1 indicates that the diffusion of lithium within the graphite particles and the diffusion of lithium in the electrolyte cannot be neglected. The estimate by Doyle et al. [9] for standard polymer batteries approximated the diffusion of lithium in the electrolyte to be three orders of magnitude larger than the solid-state diffusion of lithium ions with $S_s$ = 0.0001, thus, allowing to neglect diffusion in the active material. Our thin-layer electrode configuration suggests that the time constant of lithium diffusion in the solid contributes significantly to the overall lithiation kinetics and can therefore be investigated with the thin-layer electrode.

Table 1: Parameter set to describe the thin-layer graphite half-cells used for the experiments.

| Parameter | Value | Ref. | Parameter | Value | Ref. |
|---|---|---|---|---|---|
| $r_s$ | 1.65 µm | [10] | $\varepsilon_s = \varepsilon_e$ | 0.4 | est. |
| $D_s$ | 1·10$^{-9}$ cm$^2$ s$^{-1}$ | [3] | $h_s$ | 1 µm | meas. |
| $D_e$ | 1.4·10$^{-5}$ cm$^2$ s$^{-1}$ | [13] | $h_e$ | 200 µm | meas. |
| $c_{s,max}$ | 34175 mol m$^{-3}$ | calc. | i | 20 A m$^{-2}$ | |

To estimate the concentration gradient in the electrolyte of the half-cell, Fick's first law was approximated by a linearization at steady state conditions with an assumed transference number of the lithium ions of one:

$$J = -D_e \frac{dc}{dx} \xrightarrow[approximation]{linearized} -D_e \frac{\Delta c}{\Delta x} = J = \frac{i}{zF\varepsilon} \tag{10}$$

For the selected current density of 20 A/m$^2$ the linear approximated concentration difference would be 0.074 M across the separator with porosity ε. This value was considered negligible for a solution of 1M salt in an organic electrolyte.

The electrolyte resistivity can also be estimated using the ionic conductivity of EC:DMC 1:1 with 1M LiPF$_6$ at room temperature of 1.1 S/m [14]. For our geometry, this gives an electrolyte resistance of 3.4 Ω. Thus, the potential drop of the concentration-independent term in eq. 1 is ~9 mV when one applies the linear approximation for steady state conditions (eq. 11) which is valid for negligible concentration gradient in the electrolyte. This potential drop in the electrolyte can easily be subtracted from the total potential drop of the half-cell polarization to account for this resistive effect.

$$\frac{-i_e}{\kappa \varepsilon_e} = \nabla \phi \approx \frac{\Delta \phi}{\Delta x} \tag{11}$$

Furthermore, we can estimate the mass-transfer-limiting current density $i_{lim}$ = -zFD$_e$c$_e$(∞)/h$_e$ [15] to be 680 A/m$^2$. Considering the porosity of the separator and neglecting tortuosity, this leads to a limiting current of 36 mA, corresponding to 480C for a 0.2 mg graphite electrode. Thus, at currents higher than 36 mA, the lithium depletion in the Nernst diffusion layer will govern the response of the system in our configuration.

In summary, we come to the conclusion that the limitations in the electrolyte play a minor role in the rate performance of the thin-layer electrode up to C-rates of around 36C or even slightly higher. First, this is a result of the experimental minimization of the effects of porosity and tortuosity of a thin-layer graphite electrode. Second, the low active mass translates into a low current density in the electrolyte, causing small concentration gradients and small ohmic potential drops in the electrolyte. In addition to the negligible influence of the parameters in eqs. 1 and 2, we have used a lithium-metal counter electrode, in which diffusion of lithium within the lithium metal (eq. 5) and the ohmic potential drop in eq. 6 can be neglected except for the solid-electrolyte-interphase layer (SEI) on the lithium-metal surface.

Therefore, the equation system can be simplified. The main possible remaining influences limiting the thin-layer electrode performance are the kinetics of the surface reactions described by the Butler-Volmer equation both on the graphite working and the lithium counter electrode, the diffusion of lithium within the graphite particles, and the ohmic potential drops due to the contact resistance $R_0$ and SEI resistance. Of course, the laws of conservation of mass (eq. 2 and 5) and charge (eq. 7) also apply. Therefore, our experiments with thin-layer electrodes have mainly two side effects that influence the response of the system: the lithium counter electrode surface reactions and the contact and SEI resistances of both the lithium and the graphite electrode. Apart from this, the system response is mainly determined by the solid-state diffusion and surface reactions of the graphite particles in the thin-layer electrode. The investigation of lithium diffusion in graphite is discussed in this work while the overpotentials will be analyzed in another manuscript.

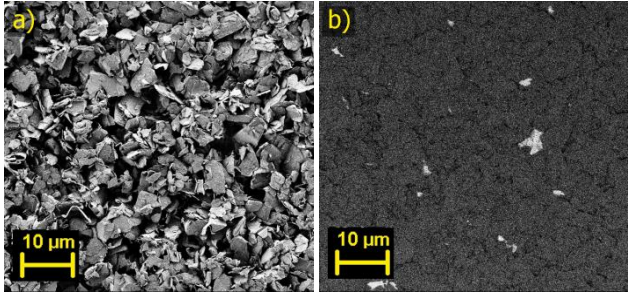

Figure 2: Scanning electron microscope images of thin-layer electrodes of SFG6 graphite; a) secondary electron detector, b) electron backscattering detector (graphite appears black and copper appears gray).

## 4. Results and discussion

### 4.1 Electrochemistry of thin-layer electrodes of graphite

Thin-layer porous electrodes of graphite SFG6 were prepared by homogeneous coating of a graphite slurry onto a copper current collector by the spray technique. The electrode is depicted in Figure 2a. The Cu current collector shines through the porous electrode in the electron back-scattering detector image in Figure 2b. From the SEM images we estimated that the porous electrode consists of 2-3 flat graphite particles stacked on top of one another. Monolayers of graphite particles can also be prepared (supplementary material, Fig. S1) but were not used for electrochemical characterization in this work because the very low active mass caused problems in the evaluation. The reasons for this, i.e., the side reaction of $Cu_xO$ layer on the copper current collector and the in-plane current distribution, will be discussed below.

To compare the thin-layer electrodes with standard graphite electrodes, we tested the rate capability of standard SFG6 electrodes (m = 4.1 ±0.05 mg, h = 55 ±2 µm). The galvanostatic responses of the standard electrodes at various rates are shown in Figure 3. The cut-off-potentials were chosen between 0.01- 1.5 V to prevent lithium plating and a possible SEI decomposition, respectively. The results confirm that standard graphite electrodes are limited to ~2C lithiation rate and 20C delithiation rate. This is in agreement with previously published results [1]. In Figure 3 one can see that the overpotential rises significantly with increasing rate for standard electrodes.

But the total overpotential decreases sharply using the thin-layer graphite electrode (m = 0.168 ±0.002 mg, h = 1 ±1 µm) as depicted in Figure 4. To compare the different kinds of electrodes, the overpotentials at 35 % SOC, i.e., during the transition from stage 2L to stage 2, were evaluated. For a lithiation rate of 2C, the overpotential of the thin-layer electrode drops from 50 mV to 11 mV and for a delithiation rate of ~20C from 640 mV to approximately 90 mV, thus, constituting less than 22 % of the overpotential of the standard graphite electrode. For comparison, we also plotted the rate capability of a graphite electrode with intermediate thickness in Figure S2 (m = 0.7 ±0.1 mg, h = 10 ±2 µm) and compared the overpotential of the stage 2L-2 transition in Figure S3. Again, a clear trend towards smaller overpotentials for thinner electrodes is observed. Note that the difference in overpotentials is obviously not caused by limitations in the graphite particles but due to diffusion limitations in the porous electrode, as discussed above.

The experiments on thin electrodes revealed that graphite particles of a median particle size of 3.3 µm along the a and b-axis [10] and ~0.4 µm along the c-axis (estimated from SEM) can be lithiated at rates up to 6C (10 min charge) with ≥82% of specific charge in the galvanostatic mode. By contrast, the delithiation can be performed at rates of up to 680C with ≥82% of specific charge delivered galvanostatically. The discharge rate capability is more than two orders of magnitude higher than the charge rate capability. This is mainly due to the fact that the applied overpotential for lithiation is limited to ~80 mV. Without this limitation, lithium plating would occur. In addition, there exist diffusion limitations within the stage 2L to 2 and stage 2 to 1 phase transitions, which decreases the respective practical specific charge associated with these transitions.

These results compare well with single-particle experiments performed on MCMB (heat-treated at 2800 °C, spherical particles of 30 µm), where a discharge rate of 1000C delivered ~85 % of the specific charge in a potential window of 0–2.5 V [16]. We can reach a similar rate in the thin-layer experiment, which confirms that our configuration is capable of investigating the properties of the active material. But one has to be careful when comparing the results for MCMB with highly crystalline flakes of graphite because MCMB is composed of very small graphitic domains mixed in between disordered carbon domains. Therefore, MCMB particles usually exhibits a higher diffusion coefficient, as shown by Takami et al. [17]. The experiment with single MCMB particles showed the characteristic plateaus of the graphite stage-transition even at a rate of 1500C, while no transition plateaus can be detected in the case of SFG6 graphite at rates higher than 170C. It seems that the crystallite size and, thus, the diffusion length are important for particle rate performance.

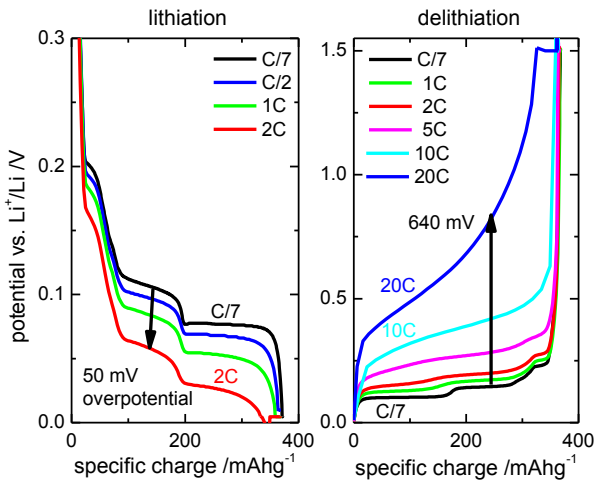

Figure 3: Electrochemical lithiation and delithiation of a typical standard graphite electrode (h = 55±2 µm) at various rates between 0.01- 1.5 V.

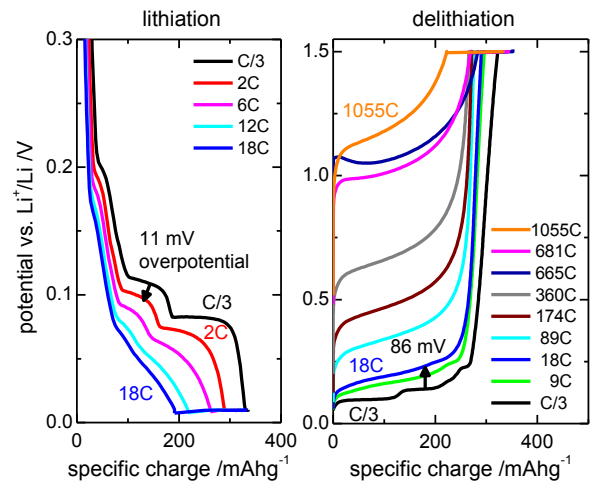

Figure 4: Electrochemical lithiation and delithiation of a typical thin-layer graphite electrode (h = 1±1 µm) at various rates between 0.01 to 1.5 V.

**4.2 Characterization of side effects**

A major challenge of the experiments with thin-layer electrodes is the influence of "parasitic reactions". The first one is the concomitant lithium insertion into the 3 wt% of SuperP conductive filler, which is needed to achieve good electronic conductivity of the electrode necessitated by the reduced number of available current pathways. Furthermore, and even more importantly, the conversion reaction of lithium with the naturally partially oxidized copper current collector needs to be considered.

Figure S4 illustrates the cycling and rate performance of SuperP electrodes, yielding 250 mAh/g of specific charge at a low rate. This translates into a parasitic charge of 8.6 mAh/g in the composite electrode. Additionally, we tested the blank copper current collector as depicted in Figure S5, for which the specific charge is normalized to a virtual graphite mass of 0.2 mg. The current collector is always slightly oxidized on the surface, exhibiting a layer of CuO on the outer surface and $Cu_2O$ between the Cu bulk and the CuO surface layer. The additional specific charge for the reduction of copper oxides was estimated to be 32 mAh/g (normalized to 0.2 mg virtual graphite mass). Based on Faraday's law and assuming $Cu_2O$ only, the calculated thickness of the $Cu_2O$ surface layer is 21 nm, which is reasonable (CuO would be 11 nm thick, see supplementary material). All other cell components (binder, titanium container, etc.) showed no measurable specific charge.

The electrochemical double-layer capacity can also be neglected. A calculation with a specific surface area of 17 m$^2$/g for SFG6, 62 m$^2$/g for SuperP, and 1.33 cm$^2$ for copper current collector (times ~10x roughness) assuming 5 µF/cm$^2$ of specific capacity gives a double-layer charge contribution of only 0.6 mAh per 0.2 mg of active material (see supplementary material). In total, we estimated that the parasitic reactions contribute more than an extra 10 % of specific charge. Especially for stage 1L and the potentiostatic step at the end of each rate cycle, the experiment would show a much higher charge than the process under consideration itself, due to the influence of the oxidic layer at the current collector (the reversible conversion reaction contributes mainly between 1.5 and 0.3 V (Fig. S5)).

In fact, there are three major challenges associated with extremely thin electrodes. They are the parasitic side reactions, determination of the correct mass of the active material and aging of the electrode during cycling. The influence of all three effects on the course of the electrochemical curves can be estimated when one superimposes the results obtained for the three different active materials (graphite, SuperP, $Cu_xO$). This means that, if the potential of the composite electrode passes through the potential window of e.g. 1.5-0.3 V in a certain period of time, one can superimpose the contribution of the three different active materials, assuming a parallel circuit that shares the total current. This assumption is only valid if no phase transitions, manifested as plateaus on the galvanostatic curves, occur within the relevant potential window, because a phase transition reaction can be considered as a "constant potential step" for the other materials.

Since electrochemical cycling was conducted at a high charge (or discharge) rate followed by a relatively moderate discharge (or charge) rate of C/3, one can precisely calculate the mass of the active material in the electrode from the course of the respective low rate half-cycle (electrochemical mass estimate). This approach is valid if the specific charge ratio of all graphite stage transitions stays constant relative to one another, allowing the identification of the loss of electronic contact of some graphite particles in the electrode. From this calculated mass we can determine the extent of particle "loss" of graphite as well as the contributions of $Cu_xO$ and SuperP for low and medium rates. At high rates, the simple linear superposition is not valid anymore due to shading effects of the graphite particles on the current collector, which increases the diffusion length within the $Cu_xO$ layer. The equation applied to estimate these effects is given by:

$$Q_{TL} = Q_{LiC_6} + {0.03}/{0.87} \cdot Q_{SuperP} + OF \cdot Q_{Cu_xO}$$
$$Q_{TL, LiC_6, SuperP, Cu_xO} = f(\Delta t, \Delta E) \tag{12}$$

with the specific charges Q of thin-layer electrode (TL), graphite (LiC$_6$), SuperP, and $Cu_xO$. The influence of SuperP has been adjusted for its mass fraction in the composite electrode and the thickness of the $Cu_xO$ layer can vary depending on the degree of oxidation, which is fitted by an oxidation factor OF.

The fitting parameters are presented and discussed in detail in the supplementary information. With this superposition approach, the active mass, aging effects, and parasitic influences can be determined with high precision. The systematic error due to these three effects can be reduced from approximately 30-36 % of the specific charge for the raw experimental data to about 6-8 %, i.e., by a factor of almost five. The linear superposition is shown in Figure S7 with the respective fractions of graphite, SuperP and $Cu_xO$. To account for stochastic errors, five nominally identical experiments have been evaluated in order to extract trends.

The determined aging of the five thin-layer electrodes is presented in Figure S6 (supplementary material). During the first three low rate cycles no significant particle loss was observed. The particle loss at the first high-rate discharge (600C) was on average 4.4% of the total mass. The next discharge at the same rate (600C) resulted in a loss of 2.5% of active mass, while the third discharge at a faster rate (1000C) resulted in an average loss of only 1.6% for the five samples. Additionally, when the same high rates were applied after more than 25 cycles, no difference to low rate aging was observed. This indicates that the stress in the material itself is not responsible for the aging. But poorly connected particles (locally less PVDF binder support) get insulated or mechanically lost in the first high-rate cycles.

It is also very important to look into the current density distribution in the thin-layer electrode. Inhomogeneous current distribution is known to occur perpendicular to the coating in standard electrodes, where the diffusion limitations of the electrolyte in the porous electrode govern the response of the system at high rates. If the thickness is very low (thin-layer electrodes), this inhomogeneity can be neglected. But inhomogeneous reactions along the lateral direction of the electrode become important. Figure 5 illustrates a monolayer and a bilayer of active particles with a corresponding equivalent circuit across the copper oxide layer and between the active particles. Assuming the previously estimated $Cu_2O$ thickness at the current collector of 21 nm and a specific resistance of 5000 Ωcm for semiconducting $Cu_2O$ [18], we can calculate the contact resistance of one particle in a monolayer (without interparticle contact). Assuming a full basal surface contact between SFG6 and $Cu_2O$, the contact resistance is calculated to be 133 kΩ which is particle size dependent (see supplementary). Although there are smaller currents for smaller particles, this would not change the electric potential of the particles relative to one another, if the thickness of all graphite particles was the same (same Q/I ratio). But this is obviously not fulfilled for real electrodes with a given particle size distribution, as seen in Figure S1 and Figure 5 for monolayers (R3 > R1 > R2), so that for monolayer electrodes no equipotential situation can be assumed.

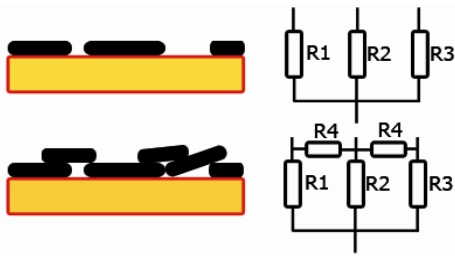

Figure 5: Sketch of lateral inhomogeneous current distribution with particles in monolayer and bilayer configuration; the copper current collector (orange) is covered with an oxide layer (red).

However, bilayers of graphite particles also possess inter-particle contacts. The good electronic conductivity of graphite and the addition of a small amount of conductive filler lead to very small resistances between the active particles (limit: R4 ≈ 0 Ω). Therefore, all single-particle resistances are in parallel to one another and an electrode with 1.33 cm$^2$ gives a calculated resistance of $14 \cdot 10^{-3}$ Ω due to the nine-million particles in parallel to each other (see supplementary material). This would be the ideal ohmic resistance assuming perfect contact between basal planes and the current collector, i.e., neglecting roughness. The real contact resistance will be in the order of a few ohms [3]. The sketch in Figure 5 shows the importance of a good inter-particle network. Therefore, the 3 wt% SuperP and a bilayer of graphite particles were chosen for the experiments.

### 4.3 Stage-dependent rate capability

The discussed galvanostatic cycling of the thin-layer electrode (as shown in Figure 4) provides more insights into the stage-dependent electrochemical performance of graphite particles. For lithiation, one can see that the 1L-4L-3L-2L stage transitions are associated with almost the same practical specific charge for 18C as for C/3 rate. By contrast, the stage 2L-2 and 2-1 transitions contribute less than half of their theoretical specific charges at a rate of 18C. Plotting these results for all five different test cells clearly demonstrates this trend (Figure 6). The formation of the graphite solid-solution (called stage 1L) consumes almost the same specific charge for a lithiation rate of 20C as for a rate of C/3. By contrast, the practical specific charge of the stage 2L-2 and 2-1 transitions decrease drastically at this high rate.

If one calculates the relative stage-specific charge fractions with respect to the low-rate charge from Figure 6, one can visualize the stage-dependent charge fraction as displayed in Figure 7, which shows the trends for the experiments more clearly. At high rates, the transitions between the liquid-like stages consume more than 80% of their theoretical specific charges, while the formation of graphite solid-solution (stage 1L) even consumes approximately 90% of its theoretical specific charge for a lithiation rate of 20C. By contrast, the transitions to the dense stages 2 and 1 consume only 40% and 45% of their maximum specific charges, respectively. This decrease is very significant because the transitions from stage 2L-2 and stage 2-1 contribute approximately 77% of the total specific charge of graphite.

These differences in behavior are not due to charge-transfer overpotentials at the electrode/electrolyte interface, because the driving forces for the transitions are comparable with a potential difference of 90 mV for the 1L-4L-3L-2L stage transition and 76 mV for the dense stage 2-1 transition. However, the differences in the stage-dependent practical specific charge at high rates can be explained by thermodynamic differences. Both stages 2 and 1 possess $LiC_6$ in-plane density [7]. By contrast, graphite stages 1L, 4L, 3L, and 2L have no in-plane order and lower in-plane densities, with suggestions varying in the literature from $LiC_9$ to $LiC_{12}$ [6, 19]. The liquid-like stages only possess ordering along the c-axis, which defines the respective stages. It seems that the solid-state diffusion coefficient of lithium in graphite is smaller for the dense stages than the liquid-like stages. This is consistent with the trends already described by Levi et al. [3] and Umeda et al. [20]. Both groups showed a difference in the lithium diffusion coefficients of one order of magnitude between the liquid-like stages compared to the dense stages.

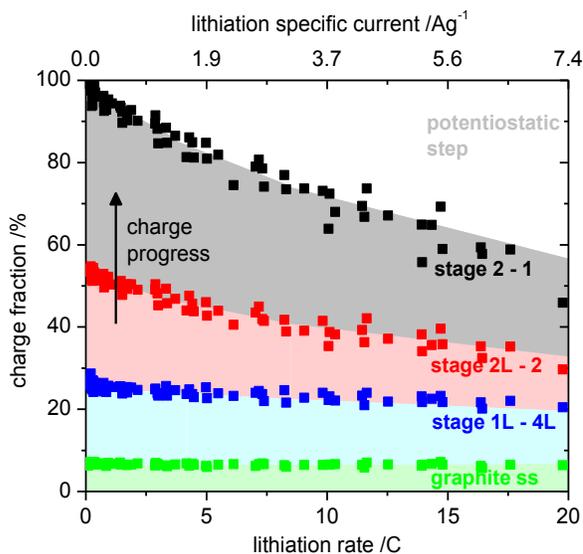

Figure 6: Charge fraction vs. lithiation rate of the different stage transitions in percent of the total specific charge (colors: green for graphite solid-solution, blue for the transition of stage 1L to 4L, red for transition of stage 2L to 2, black for transition of stage 2 to 1).

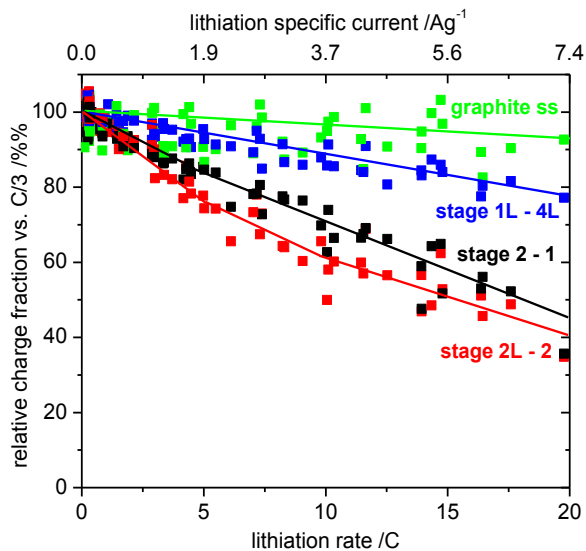

Figure 7: Relative charge fraction vs. lithiation rate for the different stage transitions, normalized to a lithiation rate of C/3.

But, interestingly, the absolute values of the diffusion coefficients in references [3] and [20] also differ by a factor of 10 to 20.

The relative contributions of the different stage transitions released by the delithiation of graphite have also been analyzed (Figure 8). Here, the same trend as for lithiation can be observed. The contributions of both stage 1 solid-solution (stage 1ss) and the transition from stage 1 to 2 decrease sharply with increasing rate. But, surprisingly, the contribution of the stage 2-2L-3L-4L transitions compensate for the initial specific charge limitation during the dense stage transitions. This can be seen more clearly in Figure 9, where each stage transition fraction is normalized with respect to its initial low rate specific charge, thus visualizing the relative changes.

Clearly, the stage transitions towards the dense stages (stage 1ss and stage 1-2) exhibit the same trend for discharge (Figure 9) as shown before for charge. Both transitions deliver only 40-45% at charge and discharge rates of 20C with respect to their low-rate specific charges. This shows that both processes, lithiation and delithiation, are symmetric with regard to the corresponding diffusion coefficient. Also, the same phase boundary movement must occur between the transitions for both charge and discharge. Here, the discussed compensation from the stage 2-2L-3L-4L transitions can be observed, delivering more practical specific charge between the rates of 1C-32C compared to the specific charge at a low rate of C/3. Of course, these stage transitions cannot deliver more than their thermodynamically defined specific charges. But it seems that the previous processes, i.e., stage 1ss and 2-1 transitions, still continue when the next stage 2-2L-3L-4L transitions start. This means that the phase boundaries of the previous stage transitions still progress further into the center of the particle while a new phase boundary develops at the edge surface of the graphite particles. Unfortunately, one cannot distinguish the plateaus of the stage 2-2L-3L-4L transitions to separate the diffusion associated with the dense stage 2 from the subsequent liquid-like phases (no detectable plateau formation). Also, the relative contributions of the following transitions from stage 4L-1L and within the stage 1L (graphite solid-solution) increase significantly compared to their very low contribution to the total charge of graphite.

In summary, we found the following for the lithiation of graphite: In addition to the low applicable overpotential, also the stage transition order is important for the comparatively low lithiation rate capability of graphite. For delithiation, however, the fast liquid-like stage transitions compensate for the slow dense stage transitions, giving them more time for diffusion, which is needed to propagate their phase boundaries further into the bulk of the particle. This can explain the long-known asymmetry in the rate capability for the charge and discharge of graphite [1].

### 4.4 Proposed diffusion pathway

To visualize this behavior, an arbitrary single particle of graphite was marked with the different stage colors using gold for stage 1, red for stage 2, and blue for stage 2L in Figure 10 which correspond to the experimentally observed colors of the pure phases due to the different band structures [8]. Nucleation of a new phase is assumed to occur at the edge plane and subsequently propagate into the center of the particle. At high rates of lithiation, the stage 2L-2 transitions cannot deliver the full specific charge (Figure 7). Therefore, as the total overpotential increases due to diffusion limitation, the surface of the particle will reach the thermodynamic potential of the next stage transition, stage 2-1. When the nucleation of the new stage 1 is initiated at the edge plane, it also propagates into the particle, taking over more and more of the edge plane. The different stages in the particle distribute in a similar manner as the annuli of trees. This is due to the fact that lithium ions can only diffuse in the interslabs of the graphene layers. Grain boundary diffusion can occur but it was shown to be five orders of magnitude slower than the in-plane diffusion of lithium-ions [21].

The phase distribution along the radial vector in Figure 10 is displayed in Figure 11. The black line indicates the phase boundary between stage 1 and stage 2, while the red one illustrates the stage 2 to 2L phase boundary. Phase boundaries in graphite are expected to be isotropic along the a and b-axis due to the hexagonal symmetry of the structure. But there is an alignment along the c-axis to minimize elastic energy, which is indicated in the figure (the calculations to this point will be presented elsewhere). One can see that, although the particle appears to be composed of more than 50% of stage 1 (the golden phase seen from the basal plane at the top in Figure 10), the actual fraction of stage 1 is only 10- 20% in the bulk of the particle, as sketched in Figure 11. This implies that one cannot judge the real state-of-charge of a single particle directly from colorimetry. Also, the Daumas-Herold domains are indicated in Figure 11 for stage 2 and 2L, where local domains of the respective stage exist but alternate after a certain island length [22]. Our annuli model is an adaptation of the shrinking core model for $LiFePO_4$ [23], which has been developed further, including information of the phase boundary alignment and using the Cahn-Hilliard phase separation model [12, 24]. Our sketch is also consistent with previous models of graphite staging based on the Cahn-Hilliard approach [22].

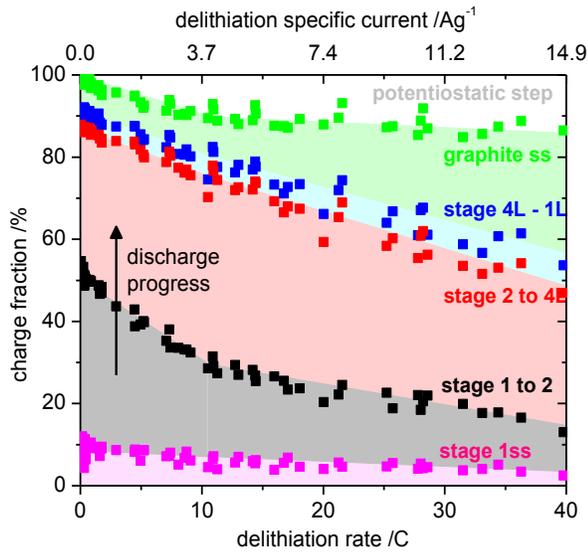

Figure 8: Charge fraction vs. delithiation rate of the different stage transitions in percent of the absolute total specific charge (colors: magenta for stage 1 solid solution, black for the transition of stage 1 to 2, red for transition of stage 2 to 4L, blue for the transition of stage 4L to 1L, green for graphite solid-solution).

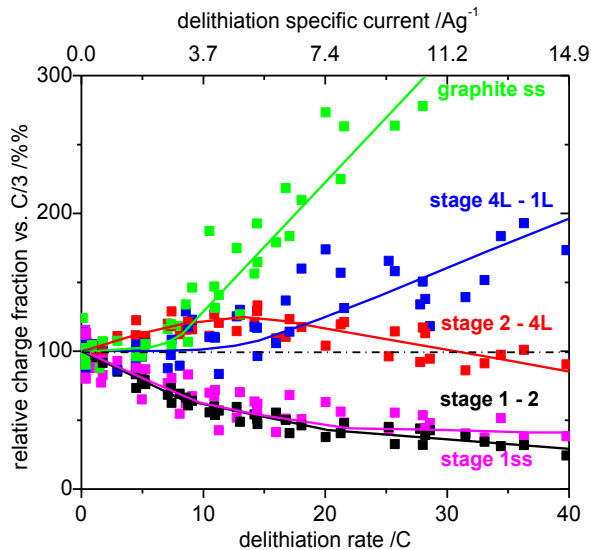

Figure 9: Relative charge fraction vs. delithiation rate for the stage transitions, normalized to the C/3 delithiation rate.

Plotting the concentration profile along the interslab third from the top in Figure 11, one can sketch the profile of the interslab concentration (Figure 12). One can see that the standard Fick's diffusion equation is not applicable over the entire particle when phase boundaries are present. Fick's law can be used in the solid-solution regime of each pure phase that develops a standard diffusion profile over the single phase regime (including many Daumas-Herold domains). But at the phase boundaries, two different diffusion mechanisms can take place. On the one hand, lithium ions could diffuse from the solid-solution phase (e.g. pure stage 1) to the interphase, which then progresses further (concentration gradient in the Daumas-Herold domains which propagates macroscopically into the particle). On the other hand, lithium ions could diffuse along the phase boundary, which was found to be the main diffusion pathway for $LiFePO_4$ [12]. At low rates, we cannot estimate the relative contribution of both diffusion pathways to the overall process, which depends on several different factors such as interfacial energy, gradient energy penalty, etc. But at high rates, diffusion limits the propagation of stages 2 and 1, as shown above. High rates lead to full coverage of the new phase at the edge plane, allowing only diffusion through the solid-solution regime of each single phase. This results in the formation of the depicted annuli.

For delithiation, a similar mechanism can be suggested. First, the stage 1 to 2 transition takes place, which is diffusion limited for the model particles with diameters of 3.3 µm. When diffusion becomes limiting in stage 2, the imposed diffusion overpotential allows the nucleation of the next phase, stage 2L, which is only 30 mV more positive than the thermodynamic potential of the stage 1 to 2 transition. Thus, the new phase, stage 2L, nucleates at the edge plane as depicted in Figure 13a. Due to the higher diffusion coefficient of this liquid-like stage 2L (as discussed above) this phase boundary progresses faster into the particle. This leads to a decrease of the stage 2 regime (Figure 13b), which increases the lithium diffusion rate due to a smaller Δr of stage 2, which in turn increases the concentration gradient dc/dr. Therefore, a higher practical specific charge can be withdrawn from the transition of stage 2 to 2L as shown above (Figure 9), because the process facilitates the stage 2 phase boundary propagation and releases its charge. The transition of stage 2 to 2L finally becomes limited by the further propagation of both stage 1 and stage 2, and therefore drops at a rate of ~32C below its low rate specific charge (Figure 9).

The next stages, 3L as well as 4L, also nucleate at the edge plane when their respective thermodynamic potential is reached. They also propagate into the particle with a high diffusion coefficient, withdrawing additional charge from all the previously uncompleted stage transitions so that almost all stored lithium ions can be released at rates of up to 680C.

At about 1000C, our experiments as well as those of Dokko et al. [16] show a major decrease in specific charge delivered galvanostatically. In our case, one knows that the current density is above the acceptable diffusion-limited current density of the electrolyte (see the theoretical considerations above). Therefore, we can conclude that graphite can be delithiated with at least 680C while at higher rates electrolyte limitations govern the response of our system. Due to the faster diffusion of lithium in the liquid-like stages, many phase boundaries can agglomerate during delithiation. But the phase boundaries are unlikely to collide, which would mean an infinitely high flux just before the collision would occur. This is due to an infinitely high concentration gradient within the pure phase dc/dr (first Fick's law). Therefore, the graphite particles are likely to simultaneously possess more annuli during delithiation.

In addition to the proposed diffusion pathway for medium and high rates there exists also an influence of the temperature on the rate capability of graphite. At higher rates, when the entire edge plane is assumed to be covered with a pure phase, only the diffusion through the solid-solution of each phase governs the kinetics. But both the diffusion coefficient and the solid-solution miscibility regime depend on temperature, as shown in the binary phase diagram of the lithium-graphite system [19]. Therefore, when the temperature is decreased the flux per area $J = -D\, dc/dr$ is reduced significantly due to a decrease in the permissible $\Delta c$ for the miscibility regime and a lower diffusion coefficient.

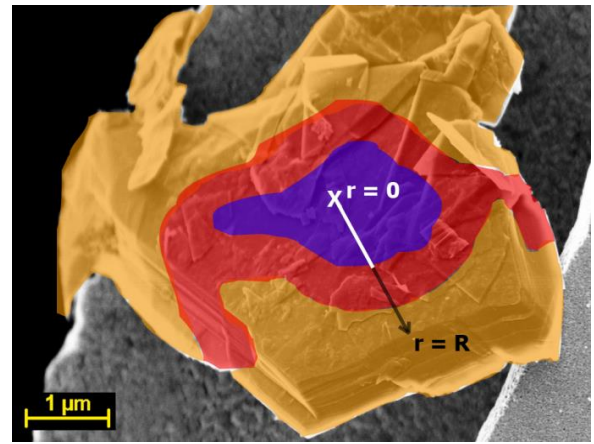

Figure 10: Shrinking annuli sketch of a single graphite particle (stage 1 in gold, stage 2 red, stage 2L blue), the arrow indicates the radial direction of the 2D cross-section shown in Figure 11.

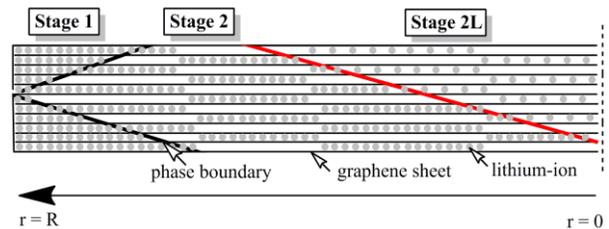

Figure 11: Sketch of a simplified 2D phase distribution profile along the arrow indicated in Figure 10 (phase boundary of stage 2L to 2 in red, phase boundary of stage 1 to 2 in black).

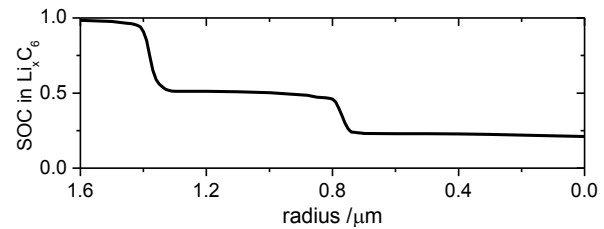

Figure 12: Sketch of the concentration profile along the third interslab from the top of Figure 11 from the edge (r = R = 1.6 µm) to the center (r = 0) in a graphite particle with phase boundaries.

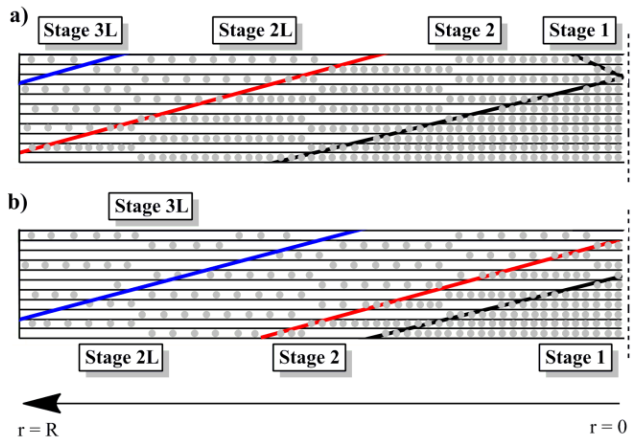

Figure 13: Simplified course of the propagation of the phase boundary during delithiation at a) the beginning and b) the end of the stage 2 to 3L transition.

## 5. Conclusions

With the newly-developed thin-layer technique, we were able to prove that graphite is a very fast lithium-ion intercalation host. Rates of 6C for lithiation and at least 680C for delithiation were reached, offering more than 80 % of the total specific charge galvanostatically. In addition to the limiting maximum overpotential of ~90 mV during lithiation, the diffusion towards the dense stages 2 and 1 also limit the rate capability of graphite. The $LiC_6$ in-plane ordered stages 2 and 1 deliver only 40-45 % of their specific charges at a lithiation rate of 20C. For delithiation, the transitions of the dense stages also limit the rate performance to the same extent as in the case of lithiation. This shows that the diffusion coefficient and the mechanism for lithiation and delithiation are similar. But surprisingly, the subsequent liquid-like stages compensate for the initial rate limitation so that almost the full specific charge can be delivered during very fast galvanostatic delithiation. From these findings, we developed a shrinking annuli model that explains the ordering of the phase boundaries. This has several implications for experiments with highly crystalline graphite electrodes (excluding MCMB and HOPG):

- The diffusion coefficient cannot be extracted by applying Fick's law over the entire particle when a phase boundary is present.
- Two different diffusion pathways are possible, i.e., the diffusion of lithium through the solid-solution regime of each pure phase (including the Daumas-Herold domains), and the diffusion along the phase boundaries between two pure phases. For high rates, the diffusion through the solid-solution regimes seems to be the main diffusion mechanism.
- The diffusion coefficient of the dense stages 1 and 2 is lower than that for the liquid-like stages, supporting the previously published trends.
- With the proposed shrinking annuli model, the asymmetry in the rate capability of graphite can be explained (in addition to the limited maximum overpotential for lithiation).

Some conclusions can also be drawn for the application of graphite electrodes in lithium-ion batteries:

- Highly crystalline standard graphite electrodes with ideally engineered porosity and electronic conductivity will probably not exceed 6C charge rate (10 min) for particles of diameters of more than 3 µm.
- A discharge at medium rates (ca. 15- 20 C) might be feasible for perfectly engineered electrodes of graphite.

Overall, the thin-layer technique is a helpful tool to study particle properties, suppressing diffusion limitations in the electrolyte. It closes the gap between single particle experiments and medium thickness electrodes and could also be used for other electrode materials.

## Supporting Information

Figures of graphite monolayer SEM, medium layer rate capability, insertion in SuperP and $Cu_xO$, aging and linear superposition are presented and described in the supplementary information. Calculations and parameter estimations are also described.

## Keywords:

Lithium-ion battery; graphite intercalation compounds; phase separation; diffusion limitations; rate capability

## Acknowledgments

The first author is grateful for the supervision by and discussions with Dr. Wolfgang Märkle (PSI) and Prof. Martin Z. Bazant (MIT). Furthermore, M.H. would like to thank the Paul-Scherrer-Institut for financial support of his PhD thesis.

## List of symbols

| | |
|---|---|
| $A$ | area, $m^2$ |
| $a_i$ | interfacial specific area, $m^2$ |
| $c_e$ | concentration of salt in electrolyte, mol $m^{-3}$ |
| $c_0$ | concentration of solvent, mol $m^{-3}$ |
| $D_e, D_s$ | diffusion coefficient in electrolyte, solid $m^2 s^{-1}$ |
| $f$ | activity coefficient, mol $m^{-3}$ |
| $F$ | Faraday's constant, 96485 C $mol^{-1}$ |
| $h$ | porous electrode thickness, m |
| $i$ | current density, A $m^{-2}$ |
| $i_0$ | exchange current density, A $m^{-2}$ |
| $I$ | superficial current density, A $m^{-2}$ |
| $J$ | diffusion flux, mol $m^{-2} s^{-1}$ |
| $m$ | mass, g |
| $OF$ | oxidation factor |
| $Q$ | charge, mAh |
| $r$ | particle radius, m |
| $R$ | universal gas constant, 8.3143 J $mol^{-1} K^{-1}$ |
| $R_0$ | contact resistance, Ω |
| $R_{SEI}$ | resistance of solid-electrolyte interphase, Ω |
| $S$ | dimensionless number for diffusion influence |
| $t^0$ | transference number |
| $T$ | temperature, K |
| $U_{OCP}$ | open-circuit potential, V |
| $z$ | number of electrons |
| $\alpha_{a,c}$ | transfer coefficient, anodic/cathodic |
| $\varepsilon$ | porosity of electrode |
| $\eta$ | surface overpotential, V |
| $\kappa$ | conductivity of electrolyte, S $m^{-1}$ |
| $\sigma$ | conductivity of solid matrix, S $m^{-1}$ |
| $\Phi$ | electrical potential, V |

Subscripts

| | |
|---|---|
| e | electrolyte in the separator |
| s | solid phase |
| n | normal direction |
| TL | thin layer |

**References**

[1] H. Buqa, D. Goers, M. Holzapfel, M.E. Spahr, P. Novak, J Electrochem Soc, 152 (2005) A474-A481.
[2] K. Persson, Y. Hinuma, Y. Meng, A. Van der Ven, G. Ceder, Physical Review B, 82 (2010) 125416.
[3] M.D. Levi, C. Wang, E. Markevich, D. Aurbach, Z. Chvoj, J Solid State Electr, 8 (2003) 40-43.
[4] J. Shim, K.A. Striebel, J Power Sources, 119 (2003) 934-937.
[5] M.D. Levi, D. Aurbach, J. Phys. Chem. B, 101 (1997) 4630-4640.
[6] J.R. Dahn, Physical review. B, Condensed matter, 44 (1991) 9170-9177.
[7] D. Billaud, E. Mcrae, A. Herold, Mater Res Bull, 14 (1979) 857-864.
[8] M.S. Dresselhaus, G. Dresselhaus, Adv. Phys., 51 (2002) 1-186.
[9] M. Doyle, T.F. Fuller, J. Newman, J Electrochem Soc, 140 (1993) 1526-1533.
[10] H. Buqa, A. Wursig, A. Goers, L.J. Hardwick, M. Holzapfel, P. Novak, F. Krumeich, M.E. Spahr, J Power Sources, 146 (2005) 134-141.
[11] J. Newman, W. Tiedemann, Aiche J, 21 (1975) 25-41.
[12] G.K. Singh, G. Ceder, M.Z. Bazant, Electrochim Acta, 53 (2008) 7599-7613.
[13] S.I. Lee, U.H. Jung, Y.S. Kim, M.H. Kim, D.J. Ahn, H.S. Chun, Korean J Chem Eng, 19 (2002) 638-644.
[14] E.J. Plichta, W.K. Behl, J Power Sources, 88 (2000) 192-196.
[15] P.B. Balbuena, Y. Wang, Lithium-Ion Batteries: Solid-Electrolyte Interphase, Imperial College Press, 2004.
[16] K. Dokko, N. Nakata, Y. Suzuki, K. Kanamura, J Phys Chem C, 114 (2010) 8646-8650.
[17] N. Takami, A. Satoh, M. Hara, I. Ohsaki, J Electrochem Soc, 142 (1995) 371-379.
[18] A.P. Young, C.M. Schwartz, J Phys Chem Solids, 30 (1969) 249-252.
[19] K.C. Woo, H. Mertwoy, J.E. Fischer, W.A. Kamitakahara, D.S. Robinson, Physical Review B, 27 (1983) 7831-7834.
[20] M. Umeda, K. Dokko, Y. Fujita, M. Mohamedi, I. Uchida, J.R. Selman, Electrochim Acta, 47 (2001) 885-890.
[21] K. Persson, V.A. Sethuraman, L.J. Hardwick, Y. Hinuma, Y.S. Meng, A. van der Ven, V. Srinivasan, R. Kostecki, G. Ceder, J Phys Chem Lett, 1 (2010) 1176-1180.
[22] P. Hawrylak, K.R. Subbaswamy, Physical review letters, 53 (1984) 2098-2101.
[23] V. Srinivasan, J. Newman, J Electrochem Soc, 151 (2004) A1517.
[24] J.W. Cahn, J Chem Phys, 30 (1959) 1121-1124.